\pdfoutput=1

\documentclass[sigconf]{acmart}

\usepackage{booktabs}
\usepackage{graphicx}
\usepackage{subfig}
\usepackage{amsmath}
\usepackage{enumerate}
\usepackage{amsfonts}
\usepackage{verbatim}
\usepackage{etoolbox}
\usepackage{multirow}
\usepackage{siunitx,tabularx,ragged2e,booktabs,caption}
\usepackage{geometry}   

\usepackage{adjustbox}
\usepackage{algorithm}
\usepackage[end]{algpseudocode}
\usepackage{verbatim}
\usepackage{mathtools}
\AtBeginDocument{%
  \providecommand\BibTeX{{%
    \normalfont B\kern-0.5em{\scshape i\kern-0.25em b}\kern-0.8em\TeX}}}

\copyrightyear{2022}
\acmYear{2022}
\setcopyright{acmlicensed}
\acmConference[WSDM '22] {Proceedings of the Fifteenth ACM International Conference on Web Search and Data Mining}{February 21--25, 2022}{Tempe, AZ, USA.}
\acmBooktitle{Proceedings of the Fifteenth ACM International Conference on Web Search and Data Mining (WSDM '22), February 21--25, 2022, Tempe, AZ, USA}
\acmPrice{15.00}
\acmISBN{978-1-4503-9132-0/22/02}
\acmDOI{10.1145/3488560.3498518}
\begin{document}
\fancyhead{}

\title{Keyword Assisted Embedded Topic Model}

%%
%% Submission ID.
%% Use this when submitting an article to a sponsored event. You'll
%% receive a unique submission ID from the organizers
%% of the event, and this ID should be used as the parameter to this command.
%%\acmSubmissionID{123-A56-BU3}

%%
%% The majority of ACM publications use numbered citations and
%% references.  The command \citestyle{authoryear} switches to the
%% "author year" style.
%%
%% If you are preparing content for an event
%% sponsored by ACM SIGGRAPH, you must use the "author year" style of
%% citations and references.
%% Uncommenting
%% the next command will enable that style.
%%\citestyle{acmauthoryear}

%%
%% end of the preamble, start of the body of the document source.

%%
%% The "title" command has an optional parameter,
%% allowing the author to define a "short title" to be used in page headers.

%%
%% The "author" command and its associated commands are used to define
%% the authors and their affiliations.
%% Of note is the shared affiliation of the first two authors, and the
%% "authornote" and "authornotemark" commands
%% used to denote shared contribution to the research.

\author{Bahareh Harandizadeh}
\affiliation{%
  \institution{Information Science Institute}
  \institution{University of Southern California}
  \city{Los Angeles}
  \state{CA}
  \country{USA}
}
\email{harandiz@usc.edu}

\author{J. Hunter Priniski}
\affiliation{%
  \institution{Department of Psychology}
  \institution{University of California, Los Angeles}
  \city{Los Angeles}
  \state{CA}
  \country{USA}
  }
\email{priniski@ucla.edu}

\author{Fred Morstatter}
\affiliation{%
  \institution{Information Science Institute}
  \institution{University of Southern California}
  \city{Los Angeles}
  \state{CA}
  \country{USA}
}
\email{morstatt@usc.edu}

%%
%% By default, the full list of authors will be used in the page
%% headers. Often, this list is too long, and will overlap
%% other information printed in the page headers. This command allows
%% the author to define a more concise list
%% of authors' names for this purpose.
\begin{comment}
\renewcommand{\shortauthors}{Harandizadeh, et al.}
\end{comment}

%%
%% The abstract is a short summary of the work to be presented in the
%% article.
\begin{abstract}
By illuminating latent structures in a corpus of text, topic models are an essential tool for categorizing, summarizing, and exploring large collections of documents. Probabilistic topic models, such as latent Dirichlet allocation (LDA), describe how words in documents are generated via a set of latent distributions called topics. Recently, the Embedded Topic Model (ETM) has extended LDA to utilize the semantic information in word embeddings to derive semantically richer topics. As LDA and its extensions are unsupervised models, they aren't defined to make efficient use of a user's prior knowledge of the domain. To this end, we propose the Keyword Assisted Embedded Topic Model (KeyETM), which equips ETM with the ability to incorporate user knowledge in the form of informative topic-level priors over the vocabulary. Using both quantitative metrics and human responses on a topic intrusion task, we demonstrate that KeyETM produces better topics than other guided, generative models in the literature\footnote{Code for this work can be found at \url{https://github.com/bahareharandizade/KeyETM}}.
\end{abstract}

%%
%% The code below is generated by the tool at http://dl.acm.org/ccs.cfm.
%% Please copy and paste the code instead of the example below.
%%
\begin{comment}
\begin{CCSXML}
<ccs2012>
<concept>
<concept_id>10002951.10003317.10003318.10003320</concept_id>
<concept_desc>Information systems~Document topic models</concept_desc>
<concept_significance>300</concept_significance>
</concept>
</ccs2012>
\end{CCSXML}

\ccsdesc[300]{Information systems~Document topic models}
\end{comment}

%%
%% Keywords. The author(s) should pick words that accurately describe
%% the work being presented. Separate the keywords with commas.

%% A "teaser" image appears between the author and affiliation
%% information and the body of the document, and typically spans the
%% page.
% \begin{comment}
% \begin{teaser}
%   \includegraphics[width=\textwidth]{sampleteaser}
%   \caption{Seattle Mariners at Spring Training, 2010.}
%   \Description{Enjoying the baseball game from the third-base
%   seats. Ichiro Suzuki preparing to bat.}
%   \label{fig:teaser}
% \end{teaserfigure}
% \end{comment}

%%
%% This command processes the author and affiliation and title
%% information and builds the first part of the formatted document.
\maketitle

\section{Introduction}
Topic models illuminate latent semantic themes within a large collection of documents. These approaches have gained mass appreciation, with applications spanning from industry to computer science research  \cite{boyd2017applications}, as well as into the humanities~\cite{schmidt2012words}. Despite their efficiency in exploring major themes of a corpus, topic models usually tend to retrieve the most general and prominent topics in the corpus, which may not necessarily align with the topics a user is interested in, and may occlude important nuance in a dataset. 

As a result, fully automated topic models (either classical or neural) are susceptible to create multiple topics with similar semantic themes that are difficult to interpret. To understand the issues better, consider retrieved topics from subset of AYLIEN COVID-19 dataset\footnote{\url{https://aylien.com/blog/free-coronavirus-news-dataset}} via LDA in Table \ref{tab:LDA}. This real-world example lays bare this important shortcoming of topic models. As is shown here, it is hard to interpret topics 1 and 4 due to coherence issues (e.g., topic 1 might related to the election or Covid). Furthermore, the topics are not diverse (e.g., topics 2 and 3 mostly refer to the same subject).

\begin{table}
  \centering
  \caption{LDA retrieved topics on part of AYLIEN COVID-19. The retrieved topics have overlapped semantically.}
  \label{tab:LDA}
  \begin{tabular}{l|l|l|l}
    \toprule
    \textbf{Topic 1} & \textbf{Topic 2} & \textbf{Topic 3} & \textbf{Topic 4}\\
    \midrule
    president & pandemic & death & democrat \\
    trump & country & covid & black \\
    coronavirus & state & coronavirus & campaign \\
    vote & coronavirus & state & elect \\

  \bottomrule
\end{tabular}
\end{table}

It is not surprising that automated topic models suffer from these setbacks, as they are not exposed to information regarding the topics of interest. In these cases, experts need to wait until the model is fit and then make post-hoc decisions about substantive topics of interest~\cite{eshima2020keyword}. Even then, topics pertaining to niche concepts are often missing.

One way to tackle the issue is to incorporate user knowledge in deriving topics by leveraging informative topic-level priors over the vocabulary. Building off the success of ETM which blends the strengths of neural topic models and word embeddings, we propose Keyword Assisted Embedded Topic Model (KeyETM), an embedded topic model guided by a user's knowledge of the corpus. The user supplies the model with a set of seed word lists associated with topics of interest, effectively placing informative topic-level priors over the vocabulary to guide statistical inference. We demonstrate the power of this approach by comparing our model to other guided topic models on the basis of quantitative measures and human responses on an intrusion task \cite{chang2009}.
\begin{figure*}[h]
  \centering
  \includegraphics[width=.7\textwidth]{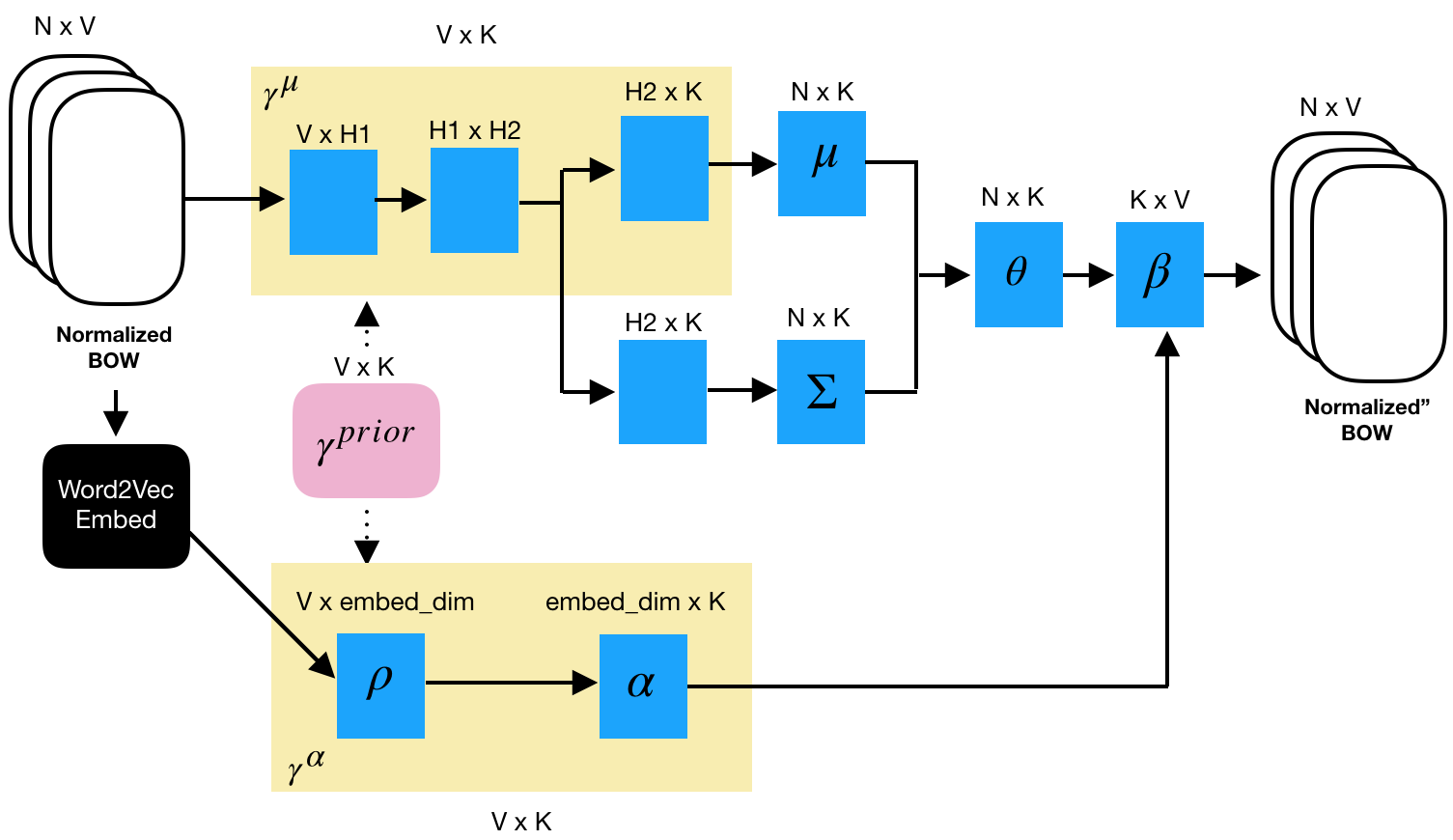}
  \caption{KeyETM Schematic: middle blue boxes represent flow of computational processing in the ETM. Remainder of figure (yellow and pink boxes), demonstrates how the model is guided by seed words. KeyETM uses document-topic distribution $\theta$ and word-topic distribution $\beta$ to estimate the marginal likelihood of a document. While $\beta$ is calculated using model parameters $\rho$ and $\alpha$ (the bottom blue boxes), $\theta$ is inferred from the variational parameters (the top blue boxes). The prior matrix (the pink box) then is defined to control these two sets of parameters independently.}
  
  \label{fig1}
\end{figure*}
Our contributions are as follows:
\begin{enumerate}
    \item Propose KeyETM, a new algorithm which equips the powerful ETM topic modeling algorithm with keyword priors.
    \item Extensively test KeyETM against several state-of-the-art baselines and show how it outperforms in terms of quantitative and qualitative topic measures.

\end{enumerate}

\section{Related Work}
\label{network}

Traditional topic models have been studied extensively. The widely-used latent Dirichlet allocation (LDA) model specifies how words are generated in documents given latent variables, or topics \cite{blei2003latent}. However, LDA suffers from limitations such as struggling to produce semantically coherent topics when fit to a large, heavy-tailed vocabulary \cite{dieng2020topic}. LDA is also computationally inefficient as it relies on Gibbs Sampling to estimate posterior densities \cite{srivastava2017autoencoding}. To overcome these issues, recent models, such as neural topic models have been suggested which employ autoencoding variational inference (e.g., Autoencoded Variational Inference For Topic Model (AVITM) \cite{srivastava2017autoencoding} or Embedded Topic Model (ETM) \cite{dieng2020topic}).

Prior knowledge of a corpus can guide models to produce better topics \cite{chen2013leveraging}. A variety of guided topic models have been proposed which turn LDA into a supervised model, such as Supervised LDA \cite{10.5555/2981562.2981578}, DiscLDA \cite{lacoste2008disclda}, and Labeled LDA \cite{ramage-etal-2009-labeled}. A critical shortcoming of these models is that they require massive collections of annotated documents to be effective. An alternative approach for guiding topic models is to supply them with topic-specific sets of seed words, operating as informative topic-level priors over the vocabulary. Popular models leveraging this form of weak supervision are Guided LDA \cite{jagarlamudi-etal-2012-incorporating} and KeyATM \cite{eshima2020keyword}). 

A similar thread of research is aspect discovery, which requires initial seed words relating to aspects of the corpus, and oftentimes can be jointly applied for sentiment analysis \cite{angelidis-lapata-2018-summarizing,huang2020weakly,hoang2019towards}. With the development of word embeddings \cite{10.5555/2999792.2999959,pennington-etal-2014-glove}, several studies have been conducted to extend topic modeling to incorporate word embeddings in deriving topics. CatE \cite{10.1145/3366423.3380278} is among these models which employs user knowledge to build a discriminative embedding space to derive topics. However, strict discrimination assumptions make them fragile when applied to a corpus with a dominating topic. Embedded Topic Model (ETM)\cite{dieng2020topic} employs variational autoencoders to learn topic posteriors and benefits from word embeddings, however, it does not incorporate user's prior knowledge in deriving topics.

\begin{figure*}[h]
  \centering
  \includegraphics[width=1.0\textwidth]{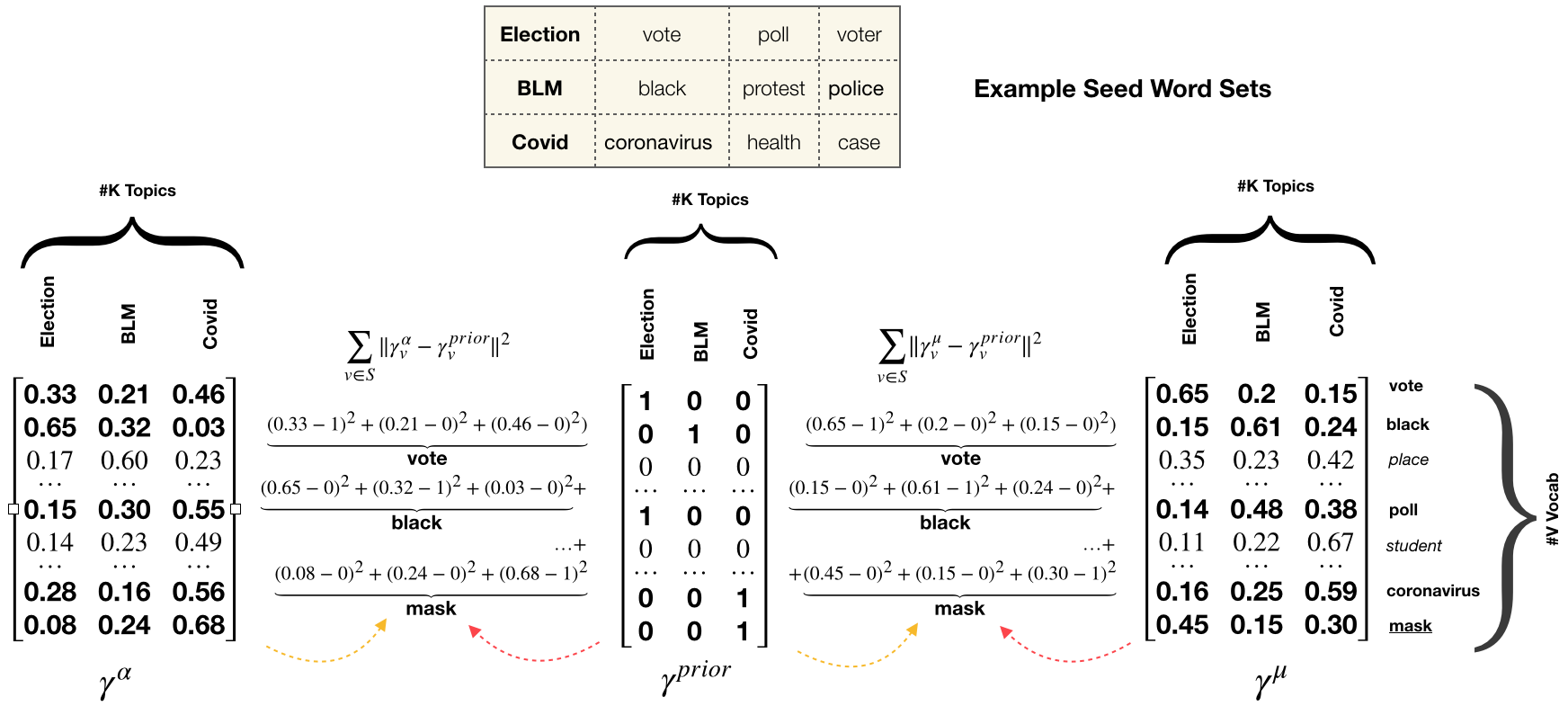}
  \caption{An example of how the square loss terms work in Equations \ref{eq.9} and \ref{eq.10}. The prior in the middle is defined based on seed word sets, then in each iteration, the algorithm will try to decrease the Euclidean distance between this prior and $\gamma^{\alpha}$(topic-word distribution), and also $\gamma^{\mu}$(topic-word distribution for documents). As a result of these changes, not only the injected words, but also the similar words to them will gradually converge to the true topics.}
  
  \label{fig2}
\end{figure*}
\section{Keyword Assisted Embedded Topic Model}
We are given a corpus of $D$ documents $\{W_{1},W_{2},....,W_{D}\}$, such that $W_{d}$ is a collection of $N_{d}$ words belonging to our vocabulary $V$. Suppose that we have a total of $K$ topics and for each topic $k$, a user provides a set of $L_k$ seed words: $\omega_{k}=\{s_{k1},s_{k2}...,s_{kL_{k}}\}$. The remainder of this section details how to fit an ETM using these keyword sets.
\subsection{Embedded Topic Model}
The ETM is a topic model that employs embedding representations of both words and topics. Therefore, the model contains two notions of latent dimensions: a word embedding matrix, $\rho \in \mathbb{R}^{V \times L}$, that embeds the vocabulary $V$ in an $L$-dimensional space (similar to the idea of classical word embeddings), and a topic embedding matrix, $\alpha \in \mathbb{R}^{K \times L}$, 
that each topic $\alpha_{k}$ (each row of the matrix) is a distributed representation of the $k^{th}$ topic in the semantic space of words. This is unlike the idea of classical topic modeling approaches such as LDA where each topic is a full distribution over the vocabulary. 
 
The \textbf{generative process} of the $d^{th}$ document within a corpus D under the ETM assumption is then defined as the following:
\begin{enumerate}
    \item Draw topic proportions: $\theta_{d}\sim \mathcal{L}\mathcal{N}(0,I)$, where $\mathcal{L}\mathcal{N}(.)$ denotes a logistic-normal distribution that transforms a standard Gaussian random variable: $\delta_{d}\sim\mathcal{N}(0,I)$ to the simplex: $\theta_{d}=softmax(\delta_{d})$. Consequently, $\delta_{d}$ and $\theta_{d}$ are called untransformed and transformed topic proportions respectively. 
    \item For each word n in the $d^{th}$ document, draw topic assignment $z_{dn}\sim Cat(\theta_{d})$, then draw the word $w_{dn}\sim softmax(\rho^{T}\alpha_{z_{dn}})$, where $\rho$ is the word embedding matrix and $\alpha_{z_{dn}}$ is the $z_{dn}$'s topic embeddings. The $Cat(.)$ also denotes the categorical distribution. 
\end{enumerate}

\noindent The \textbf{inference process} in ETM is the estimation of its two sets of model parameters: word embeddings $\rho_{1:V}$ and topic embeddings $\alpha_{1:K}$. To achieve the goal, the marginal likelihood of documents is maximized: 
\begin{equation}
\label{eq.1}
\mathcal{L}(\alpha,\rho) = \sum_{d=1}^{D}\log p(W_d|\rho,\alpha).
\end{equation}

\noindent Under ETM's generative model, the marginal likelihood of a document $W_d$, after marginalizing out each word topic assignment $z_{dn}$, is equal to:
\begin{equation}
\label{eq.2}
p(W_d|\rho,\alpha) = \int p(\delta_{d}) \prod_{d=1}^{N_{d}}p(w_{dn}|\delta_{d},\alpha,\rho)d\delta_{d},
\end{equation}

\noindent where the distribution of $w_{dn}|\delta_{d},\rho,\alpha$ can be estimated as:

\begin{equation}
\label{eq.3}
p(w_{dn}|\delta_{d},\rho,\alpha) = \sum_{k=1}^{K}\theta_{dk}\beta_{k,w_{dn}}.
\end{equation}

\noindent So one only needs to evaluate the distributions of $\theta$ and $\beta$ to calculate this probability. Similar to LDA, $\theta_{dk}$ in Equation \ref{eq.3} denotes topic proportions for each document, and $\beta_{kv}$ shows a topic distribution over vocabulary ($V$), but unlike LDA, $\theta$ is calculated based on step 1 in ETM's generative process and $\beta$, inspired by the continuous bag-of-words (CBOW) \cite{DBLP:journals/corr/abs-1301-3781} model, is induced using: $softmax(\rho^T\alpha_{k})|_{v}$ (see step 2 in ETM's generative process). 

Finally, Equations \ref{eq.3} and \ref{eq.2} can flesh out the likelihood in Equation \ref{eq.1} and the solution seems to be obtained completely. However, the calculation of marginal likelihood of each document in Equation \ref{eq.2} is hard to compute due to the existence of an intractable integral over the topic proportions. To tackle the issue, ETM uses the evidence lower bound (ELBO) function that includes both the model parameters $(\rho,\alpha)$, and the variational parameters $(\nu)$:
\begin{equation}
\label{eq.4}
\begin{aligned}
  \mathcal{L}(\Theta) =  \sum_{d=1}^{D}\sum_{n=1}^{N_{d}} \mathbb{E}_{q}[\log p(w_{dn}|\delta_{d},\rho,\alpha)] - \\
  \sum_{d=1}^{D} KL(q(\delta_{d};w_{d},\nu) || p(\delta_{d})),
  \end{aligned}
\end{equation}

\noindent where $\Theta$ represents the set of all the model and variational parameters and $q(\cdot)$ is a Gaussian whose mean and variance come from a neural network, parameterized by variational parameters ($NN(x;\nu)$). 

\noindent As a function of the model parameters, the objective function in Equation \ref{eq.4} tries to maximize the expected complete log-likelihood. As a function of the variational parameters, the first term pushes parameters to place mass on topic proportions, which explains the observed words better, and the second term forces them to be as close as possible to the prior. For more information, see the blue boxes in Figure \ref{fig1}, the top blue boxes show how $\theta$ is obtained and the bottom blue boxes show $\beta$ calculation. 

% FM Comment: i removed this because it is sort of redundant. it's also not good to mention "substantial improvement" before we demonstrate it..
%Although ETM achieves substantial improvement compare to many topic models such as LDA, it is still susceptible to describing the most dominant themes of documents. As a result, injecting a users' prior knowledge can guide the model towards themes that would not otherwise emerge. 
Now we describe how ETM can be modified to utilize a user's prior knowledge about the corpus to infer desired topics.

\subsection{KeyETM Description}

%The evidence lower bound(ELBO) function in Eq.\ref{eq.1},  is optimized with respect to both the model parameters $(\rho,\alpha)$ and the variational parameters$(\nu)$.

KeyETM extends ETM to incorporate user-defined seed words ($\omega_{1:K}$) when inferring topics. First, since ETM with pre-fit word embeddings (i.e., labeled ETM) is considered as the base model, $\rho$ is trained on the given corpus $D$ using skip-gram embeddings \cite{mikolov2013distributed} and is considered fixed during KeyETM training. Then, for each topic $k$, its corresponding semantic vector $\omega_{ks}$ is defined as a mean of all word embedding vectors in $\omega_{k}$:
\begin{equation}
\label{eq.5}
\omega_{ks} = \frac{\sum_{i=1}^{L_{k}}\rho(s_{ki})}{L_{k}}
\end{equation}

\noindent and based on that, each element of the prior knowledge matrix $\gamma^{prior} \in \mathbb{R}^{V \times K}$ is defined as follows:
 \begin{equation}
    \label{eq.6}
    \gamma_{vk}^{prior}=
    \begin{cases}
      1, & \text{if}\ ((v\in \omega_{k}) \parallel (CosSim(\rho(v),\omega_{ks}) \geq thr)) \\
      0, & \text{else}.
    \end{cases}
  \end{equation}

\noindent In Equation \ref{eq.6}, for all words $v \in V$ and in each topic $k$, if $v$ is a member of topic $k$'s seed words ($v \in \omega_{k}$) or $v$ not in topic $k$'s seed words, but its similarity to topic $k$'s semantic vector ($\omega_{ks}$) is greater than a threshold, the prior value is set to one. Otherwise, the prior is assigned to zero ($\gamma^{prior}$ is shown with the pink box in Figure \ref{fig1}). 

Using matrix $\gamma^{prior}$, set $S$ is defined as a union of all words with at least one value greater than zero in their corresponding rows in $\gamma_{prior}$ matrix:
\begin{equation}
\nonumber
    S = \{v | \exists \ k, \gamma_{vk}^{prior}=1\}
\end{equation}

\noindent In other words, $S$ is a union of all seed words for all topics plus similar words to them.

After the definition of primitive variables, the ETM's objective function needs to be changed. As mentioned in section 3.1, the ELBO function in Equation \ref{eq.4} is optimized with respect to both the model parameters and variational parameters. Consequently, we need to penalize both sets of parameters to force the model to infer the desired topics. 

To penalize variational parameters, inspired by AVIAD model\cite{hoang2019towards}, $\gamma^{\mu} \in \mathbb{R}^{V \times K}$ is defined as the topic distribution over vocabularies $V$ corresponding to document $d$. $\gamma^\mu$ is shown in the top yellow box in Figure \ref{fig1}. As you see in this box, we have three linear layers: let $W_{1} \in \mathbb{R}^{H_{1} \times V}$ and $b_{1} \in \mathbb{R}^{H_{1}}$ be the parameters in the first layer, $W_{2} \in \mathbb{R}^{H_{2} \times H_{1}}$ and $b_{2} \in \mathbb{R}^{H_{2}}$ be the parameters in the second layer, and $W_{3} \in \mathbb{R}^{H_{2} \times K}$ and $b_{3} \in \mathbb{R}^{K}$ be the parameters for the third one. To estimate $\gamma^{\mu}$ for each document, we first calculate the following steps:
\begin{equation}
 \label{eq.7}
\begin{split}
    M =SoftPlus(W_{1}^T+b_{1}) \\
    N = SoftPlus(M.W_{2}^T+b_{2}) \\
    W_{dr} =  dropout(N,dropout\_rate) \\
    W_{mean} =  Softmax(W_{dr}.W_{3}^T+b_{3}), 
    \end{split}
\end{equation}

\noindent where $W_{mean} \in \mathbb{R}^{V \times K}$. Then, each element of $W_{mean}$ is multiplied  with scalar value $0$ if the word corresponding to that row is not available in document $d$ and $1$ otherwise.

To penalize model parameters, as word embedding $\rho$ is considered fixed, we need to force topic embedding $\alpha$ to be as close as possible to the desired topics. Therefore, $\gamma^\alpha$ is defined as the output of this transformation: 
\begin{equation}
 \label{eq.8}
\gamma^\alpha = Softmax(\rho.\alpha^{T}),
\end{equation}
which is shown using the bottom yellow box in Figure \ref{fig1}. In fact, $\gamma^\alpha$ is equal to $\beta$ and shows the distribution of topics over words, but we choose another name to make it more consistent.

\noindent By using the prior matrix $\gamma_{prior}$, the variational regularization term $L_{\mu}$ is defined as:
\begin{equation}
 \label{eq.9}
    L_{\mu}=\sum_{d}\sum_{v \in S}\lVert \gamma_{dv}^{\mu}-\gamma_{v}^{prior} \lVert^2
\end{equation}
and the model regularization term is defined as: 
\begin{equation}
 \label{eq.10}
L_{\alpha}=\sum_{v \in S}\lVert \gamma_{v}^{\alpha}-\gamma_{v}^{prior} \lVert^2
\end{equation}

\noindent Both losses are just applied for each word $v$ that exist in set S. The square loss in Equation \ref{eq.9} will penalize the variational parameters to make the inference neural network produce the distribution $\gamma^\mu$ as similar to the prior distribution $\gamma^{prior}$ as possible, and the square loss in Equation \ref{eq.10} penalizes the topic embedding $\alpha$ to produce topics near the selected themes. 

\noindent Finally, the loss function in Equation \ref{eq.4} is modified by adding these two new regularization terms: 
 
 \begin{equation}
 \label{eq.11}
   \mathcal{L}(\Theta) =
    ELBO - \lambda_{1} L_{\mu}- \lambda_{2}L_{\alpha}
  \end{equation}

To better demonstrate the process, consider the example that is shown in Figure \ref{fig2}. Given the three topics, namely \textit{Election}, \textit{BLM} and \textit{Covid}, we chose three seed words for each topic and showed them in the top middle table. Then, $\gamma^{prior}$ was constructed based on this table, such that each element of the matrix $\gamma^{prior}_{vk}$ is set to 1 if $v \in \omega_{k}$ and 0 otherwise (e.g., the word \textit{``vote''} is received prior 1 for \textit{Election} topic and 0 for the rest). Moreover, the similar words to topic k's semantic vectors $\omega_{ks}$, also receive the prior of 1 for the corresponding topic (e.g., \textit{mask} for topic \textit{Covid}).

\noindent Assume at the $k^{th}$ iteration of the training process, the general topic-word distribution $\gamma^{\alpha}$ is in the state that is displayed in the left matrix and $\gamma^{\mu}$ is the topic distribution for each word corresponding to a document $d$ that is shown in the right matrix. Then minimizing the Euclidean distance between $\gamma^{\alpha}$ and $\gamma^{prior}$ for $v \in S$ (e.g., \textit{vote},\textit{black},\textit{poll}, etc.) encourages topic embedding $\alpha_{1:K}$ gradually generate topics near to the selected themes, and minimizing the Euclidean distance between $\gamma^{\mu}$ and $\gamma^{prior}$ forces the variational parameters to place mass on topic proportions that explain the injected words and the similar words to them better. As a result of these changes, not only do the seed words receive high topic posteriors, but other related words define the topic as well. In effect, shedding light on the broader topics in which the seed words are related to. The complete procedure is shown in Algorithm 1, when $NN(x;\nu)$ is used to represent a neural network with input x and parameters $\nu$.

\begin{algorithm}
\caption{KeyETM topic modeling}
\begin{algorithmic}[1]
\State Choose K and define $\omega_{1:K}$
\State Create word2vec embedding from dataset and define $\rho$
\State Define the semantic vector$\omega_{ks}$, for $\omega_{1:K}$ based on $\rho$ 
\State Initialize prior matrix $\gamma^{prior}$ using $\omega_{1:K}$ and $\omega_{(1:K)s}$
\State Initialize model and variational parameters
\For{iteration i=1,2,\ldots }
    \State Compute $\beta_{k}=$softmax$(\rho^{T}\alpha_{k})$ for each topic k
    \State Choose a minibatch $N$ of documents
    \For{ each document d in $N$ }
        \State Get normalized bag-of-word $x_{d}$
        \State Compute $\mu_{d}=NN(x_{d};\nu_{\mu})$
        \State Compute $\Sigma_{d}=NN(x_{d};\nu_{\Sigma})$
        \State Sample $\theta_{d} \sim \mathcal{L}\mathcal{N}(\mu_{d},\Sigma_{d})$
        \For{each word in the document d}
            \State Compute $p(w_{dn}|\theta_{d})=\theta^{T}_{d} \beta_{.,w_{dn}}$
        \EndFor
        \EndFor
        \State set $\gamma^{\alpha} \leftarrow \beta$
        \State Compute $L_{\mu} = \sum_{d}^{N}\sum_{v \in S}\lVert \gamma_{dv}^{\mu}-\gamma_{v}^{prior} \lVert^2$
        \State Compute $L_{\alpha} = \sum_{v \in S}\lVert \gamma_{v}^{\alpha}-\gamma_{v}^{prior} \lVert^2$
        \State Estimate ELBO ($kl_{\theta}$ + reconstruction loss)
        \State Estimate the total loss:  $\mathbb{L}\leftarrow$ ELBO - $\lambda_{1}L_{\mu}$ - $\lambda_{2}L_{\alpha}$
        \State Estimate gradient (backprop.) 
        \State Update model parameters
    
\EndFor

\end{algorithmic}
\end{algorithm}
\section{Evaluation of KeyETM}
In this section, we first discuss the role of some hyper-parameters in KeyETM, then the model's performance will be studied and compared to other guided topic models. In particular, topics coherency and diversity will be examined by designing quantitative and qualitative experiments, and classification performance will be reported to analyze document representations.  

\begin{comment}
A good topic model in general should provide both coherent patterns of language and diversity. In other words, the inferred topics not only should be interpretable, but also include a range of aspects. 
\end{comment}

\begin{comment}
In quantitative results, we reported topic interpretability as a blend of topic coherence and diversity\cite{dieng2020topic}. We also employed an intrusion task~\cite{chang2009} to assess how human raters evaluated the coherence of terms inside each topic. In qualitative section, we reported top words for all topics to examine how these words are related to the desired topics.
\end{comment}
 
\begin{comment}
\begin{table}
  \centering
  \caption{User-Defined Topics and Keywords}
  \label{tab:keywords}
  \begin{tabular}{ll}
    \toprule
    Topic & Keywords\\
    \midrule
    \textbf{Election} & elect, voter, vote, campaign, poll\\
    \textbf{Economy} & econom, economi, busi, job, market\\
    \textbf{General Covid-19} & coronaviru, health, case, covid, diseas \\
    \textbf{Pandemic Response} & mask, governor, wear, face, order\\
    \textbf{Black Lives Matter} & black, protest, polic, racist, ralli \\
  \bottomrule
\end{tabular}
\end{table}
\end{comment}

\noindent \textbf{Dataset:} We used the AYLIEN COVID-19 dataset\footnote{\url{https://aylien.com/blog/free-coronavirus-news-dataset}} which consists of over 1.5M news articles about the Covid-19 pandemic, with articles spanning from November 2019 to July 2020. We then limited the articles to the ones that were published in the Huffington Post, CNN, Breitbart and, Fox News. To ease model fitting, we extracted a subset of $9,663$ posts which had one of the following IPTC Subject Codes \footnote{\url{https://docs.aylien.com/newsapi/search-taxonomies/\#search-labels-for-iptc-subject-codes}}: \textit{medical procedure}, \textit{disease}, \textit{election}, \textit{social issues}, \textit{unemployment}, \textit{employment}, \textit{media}, \textit{business} and \textit{economy}. This selection criteria resulted in a corpus with imbalanced topic representation. For instance, there are nearly $3,200$ articles discussing \textit{disease} and $457$ about \textit{racism}. Also as the name of the dataset suggests, Covid is a dominant theme, which implies, in addition to documents that talk about Covid directly (e.g., disease), other documents may also relate to an aspect of Covid (e.g., election or employment). This imbalance captures the ecological conditions in which practitioners often apply topic models, and we believe provides a verdant assessment of how the evaluated models will fair in the field. 

\noindent To have a more balanced dataset, without a dominant theme, the 20Newsgroups corpus was also used which is a collection of newsgroup posts. We extracted a subset of 6,000 articles from \textit{religion}, \textit{politics} and \textit{science} subjects.     
Both datasets were preprocessed by filtering stop words, and words with document frequency above 75\% and below 10\%. We fit topic models to the stems summaries of the articles using PorterStemer\footnote{\url{https://www.nltk.org/howto/stem.html}}. 
\begin{comment}
\begin{figure}[h]
  \centering
  \includegraphics[width=.5\textwidth]{samples/keyword.pdf}
  \caption{Seed word sets and their proportion. Proportion is defined as a number of times a keyword occurs in the corpus divided by the total length of documents.}
  \label{fig3}
\end{figure}
\end{comment}

\begin{figure}%
    \centering
    \subfloat{{\includegraphics[width=.55\textwidth]{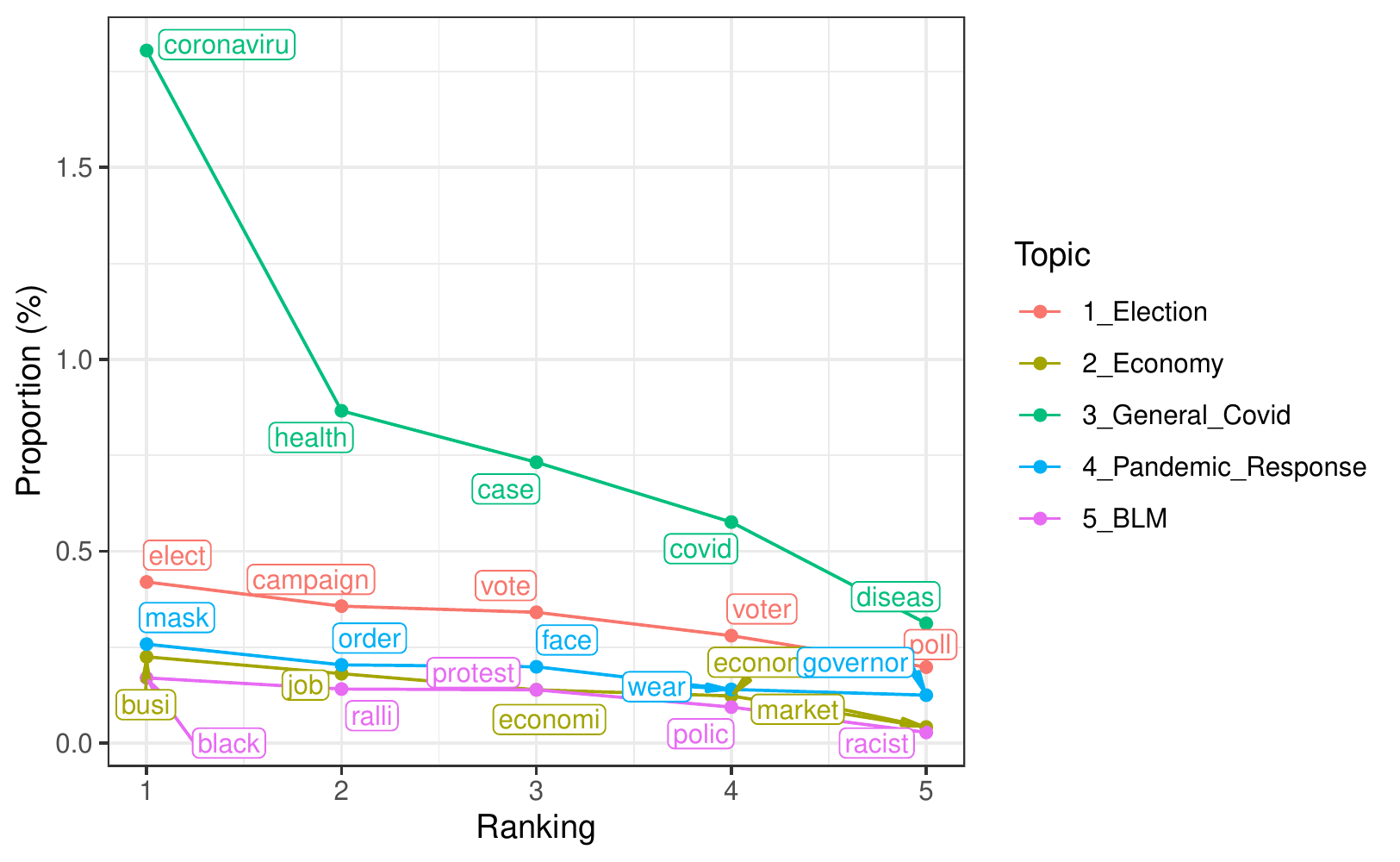} }}%
    \qquad
    \subfloat{{\includegraphics[width=.5\textwidth]{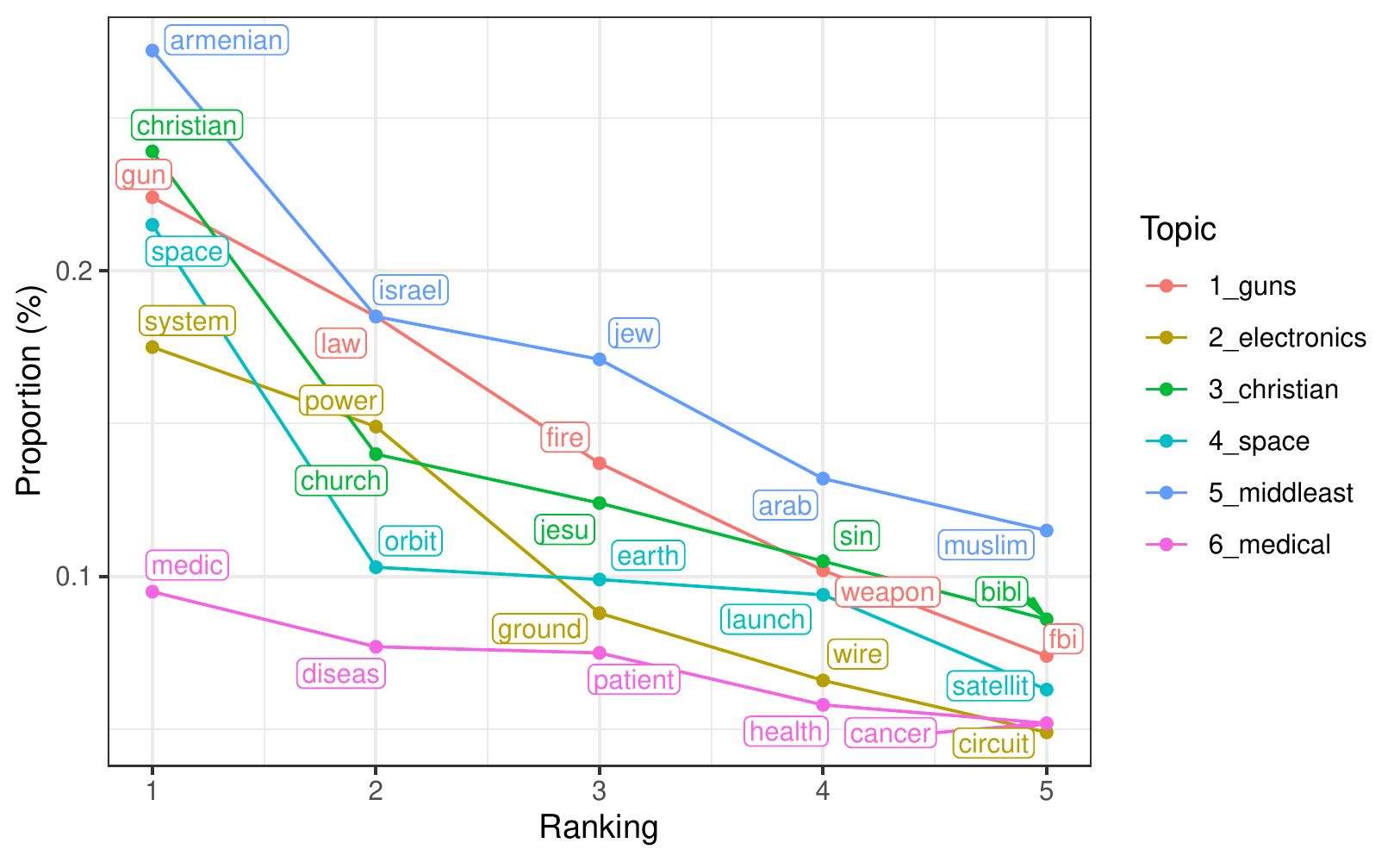} }}%
    \caption{Seed word sets and their proportion. Proportion is defined as a number of times a keyword occurs in the corpus divided by the total length of documents. The top figure shows AYLIEN COVID-19 and the bottom one shows 20Newsgroups.}%
    \label{fig:3}%
\end{figure}

\noindent\textbf{Seed words:} The seed word selection process is performed iteratively by human experts and includes running unguided topic models such as ETM and examining the relative frequency of candidate keywords. However, due to the ability of the KeyETM model to discover similar words from the pre-defined seed words (Equation \ref{eq.6}), there is no need to put so much effort to complete this step. 

\noindent  For AYLIEN COVID-19 dataset, we defined $5$ sets of keywords pertaining to different issues during the Covid-19 pandemic and for 20Newsgroups, we considered 6 sets of keywords, each captures one aspect of the dataset. As we used the stems summaries of the articles to feed the models, it is essential to also use stems for seed words definition. Fig~\ref{fig2} communicates them and their proportions in the dataset.

\noindent As it shows in Figure \ref{fig2}, the seed words proportions highlight the imbalanced representation of topics in the AYLIEN COVID-19 (e.g., the rate of terms related to Covid-19 dwarfs the rate of terms about the Black Lives Matter movement), while these proportions seem more balanced for 20Newsgroups and only medical documents are less representative.  Guided topic models should be able to effectively recover dwarfed topics given when provided with the proper set of keywords.

\subsection{Finding a Balance Point}
As it is shown in Section 3, KeyETM uses ELBO function to approximate a posterior over the document representation $\theta_{d}$ and the model parameter $\beta$; while $\beta$ is calculated using model parameters ($\rho,\alpha$), $\theta$ is inferred from the variational ones. Two additional regularization terms  ($L_{\mu}$,$L_{\alpha}$) are then applied to the ELBO to control $\theta$ and $\beta$ independently (see Equation \ref{eq.11}). 

\begin{figure}
  \centering
  \includegraphics[width=.47\textwidth,height=0.40\textheight]{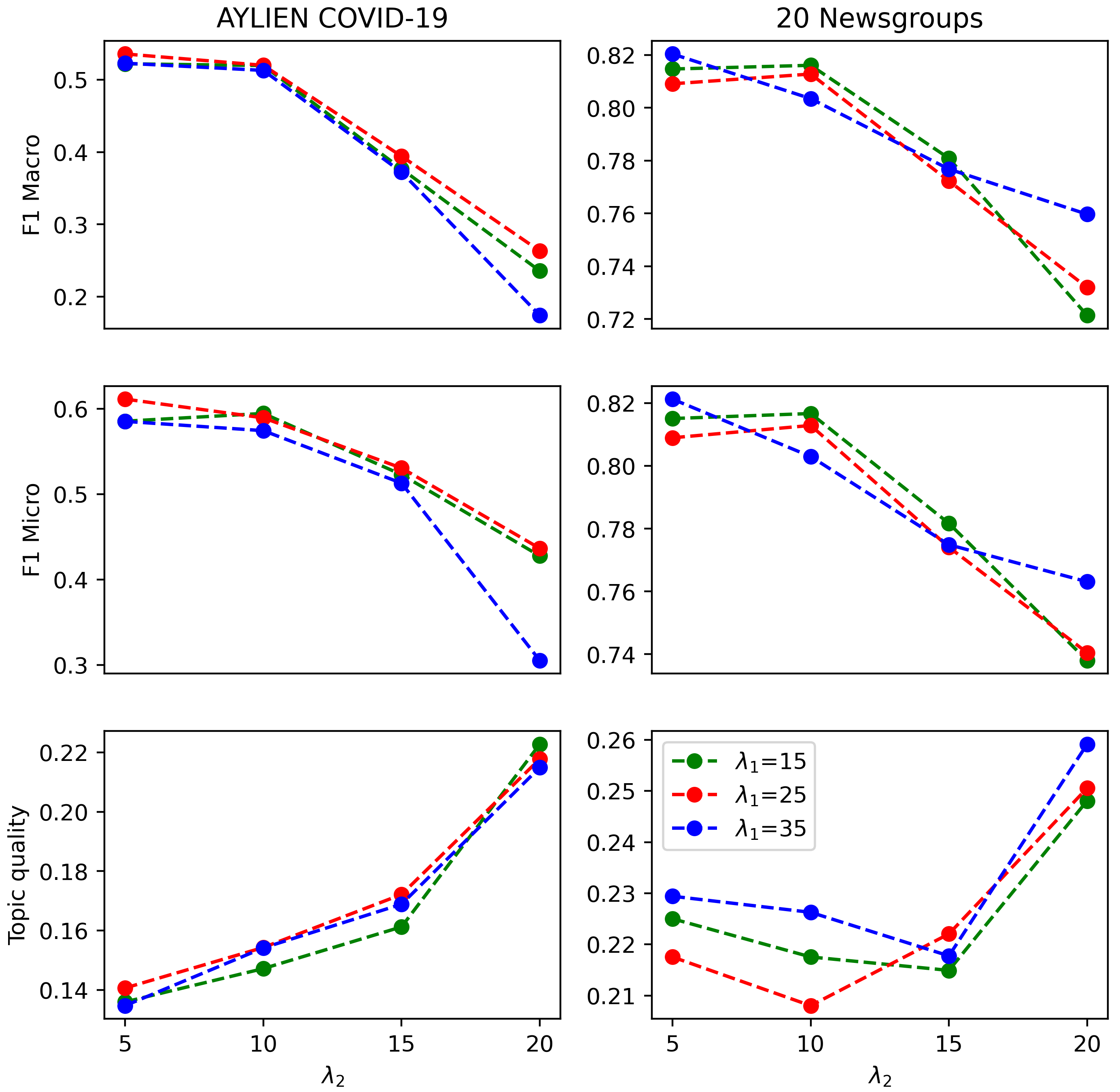}
  \caption{Document representation versus topic quality, when we change $\lambda_{1}$(weight of variational regularization) and $\lambda_{2}$ (weight of model regularization)}
  \label{fig4}
\end{figure}

One important question that is left to answer is how these additional losses work with respect to each other and how they will effect the final outputs. To address this, in Figure \ref{fig4}, for a fixed value of $\lambda_{1}$ (weight of $L_{\mu}$), that is identified with the different colors in each subplot, we change the value of $\lambda_{2}$ (weight of $L_{\alpha}$) to see their effects on F1 score and Topic Quality. Here, F1 is considered as a index of a document representation quality($\theta_{d}$), and the Topics Quality\cite{dieng2020topic} is used to analyze $
\beta$ (to get more information about Topic Quality see section 4.4). 

As you see in Figure \ref{fig4}, while the $\lambda_{2}$ is increased, at some point, the F1 decreases but the Topic Quality increases. The reason is, when we put so much weight on model regularization ($L_{\alpha}$), $\gamma^{\alpha}$ will gradually converge to prior matrix, consequently, a higher quality in topic-word distribution matrix ($\beta$) is obtained. However, as the weight for variational regularization ($L_{\mu}$) is fixed, the model's reconstruction loss from ELBO will dominate the changes of the variational parameters (to minimize the distance between input and output), and as a result, the documents representation ($\theta$) will shift towards the dominant topics. The more the dataset is unbalanced or having a dominant theme, the more stronger this effect will be emerged.

\begin{table*}
  \caption{Resulting Topics for Four Guided Topic Models}
  \label{tab:topwords}
  \begin{tabular}{ccccc|}
    
    \multicolumn{5}{c}{\textbf{Guided LDA}}\\
    \midrule
    fraud & foundat & websit(\textbf{x}) & face & virtual(\textbf{x})\\
    voter & contest & move(\textbf{x})&weekend(\textbf{x}) &appoint(\textbf{x})\\
    speak(\textbf{x}) & payment & recov & advic & patient(\textbf{x}) \\
    year(x) & speak(\textbf{x}) & respect(x) & describ(\textbf{x}) & black\\
    maintain(\textbf{x}) & encourag(\textbf{x}) & therapi & constitut(\textbf{x}) & suspect\\
  \bottomrule
\end{tabular}
  \label{tab:freq}
  \begin{tabular}{ccccc}
    
    \multicolumn{5}{c}{\textbf{AVIAD}}\\
    \midrule
    suprem & paid & respiratori& nurs&minneapoli\\
    contest(\textbf{x}) & labor & symptom &provinc&respect\\
    ballot & payment & nurs &respiratori(\textbf{x})&polit \\
    gop & financi & provinc(\textbf{x}) &wash &msnbc(\textbf{x})\\
    initi(x) & econom & europ(\textbf{x}) & claim(x) &tim(\textbf{x})\\
  \bottomrule
\end{tabular}
    \label{tab:freq}
  \begin{tabular}{ccccc|}
    
    \multicolumn{5}{c}{\textbf{KeyATM Base}}\\
    \midrule
    biden & worker & coronaviru& mask&trump\\
    trump & coronaviru(\textbf{x}) & health &state&presid(\textbf{x})\\
    elect & busi & nurs &coronaviru&coronaviru(\textbf{x}) \\
    vote & unemploy & peopl(\textbf{x}) &wear &white(\textbf{x})\\
    democrat & work & viru & order &hous(\textbf{x})\\
  \bottomrule
\end{tabular}
  \label{tab:freq}
  \begin{tabular}{ccccc}
    
    \multicolumn{5}{c}{\textbf{KeyETM}}\\
    \midrule
    campaign & job & coronaviru& trump(\textbf{x})&black\\
    elect & unemploy & health &presid(\textbf{x})&polic\\
    democrat & economi & case & order &ralli \\
    vote & econom & covid &viru &protest\\
    voter & financi & death & mask &floyd\\
  \bottomrule
\end{tabular}
  \begin{tabular}{ccccc}
    
    \multicolumn{5}{c}{\textbf{ETM}}\\
    \midrule
      trump & state & worker& peopl&coronaviru\\
    presid & elect & work &coronaviru&health\\
    biden & vote & pandem &mask&case \\
    campaign & voter & week &live &viru\\
    news & democrat & job & black &covid\\
  \bottomrule
\end{tabular}
\end{table*}

\noindent For instance, in AYLIEN COVID-19 (shown in the left side of Figure \ref{fig4}) with a dominant theme in all documents (Covid) and also an imbalance issue, the F1 scores drop almost 60\%, when we change $\lambda_{2}$ from 5 to 20, while the Topic Quality is increased from 0.14 to 0.22. Whereas, in the 20Newsgroups dataset without dominant theme and less unbalancing issue, the F1 drop less than 10\% and the topic quality also is changed in the shorter range. The reason of sudden drop in F1 for the AYLIEN dataset is that the documents representations $\theta$, is converged to the dominant topics (topic 3 and 4), while in 20Newsgroups we just observe less accurate results for the less representative topic (medical).   

As a consequence, we need to find a balance point between Topic Quality and document representation (specifically when we are using unbalanced dataset), and adjust these two parameters based on our needs to avoid the model overfitting.  
\subsection{Baseline Methods}
We compared KeyETM to three other key-words guided models (GuidedLDA \cite{jagarlamudi-etal-2012-incorporating}, AVIAD \cite{hoang2019towards}, and KeyATM \cite{eshima2020keyword}) using identical sets of seed words we listed in Figure \ref{fig:3}. Like KeyETM, all these models are generative, making them easy for comparison.  

\textbf{GuidedLDA}\footnote{ \url{https://guidedlda.readthedocs.io/en/latest/}} implements latent Dirichlet allocation (LDA) by placing topic-level priors over vocabulary informed by the seed words. Seed\_confidence is the parameter that controls how much extra boost should be given to a word and it can be range between 0 and 1. We set this parameter to 0.6 and number of epochs to 250.  
    
\textbf{AVIAD} \cite{hoang2019towards} extends the Autoencoding Variational Inference for Topic Models (AVITM) approach \cite{srivastava2017autoencoding} to embed prior knowledge by rewriting loss function to infer desired topics. It uses a variational autoencoder with the reparameterization trick to simulate the sampling task, while it does not consider word embedding. To train the model, we set learning rate to 0.001, Dirichlet parameter $\alpha=1.0$, $\lambda=20.0$, hidden\_size in encoder 1 and 2 to 150 and run the model for 150 epochs. 

\textbf{KeyATM(Base)}\footnote{ \url{https://keyatm.github.io/keyATM/}} extends GuidedLDA to use term weights to penalize highly frequent words in the corpus, to avoid dominate the resulting topics, but like GuidedLDA uses collapsed Gibbs sampling to sample from the posterior distribution. The model allows topics to have empty seed words sets, which are called no\_keywords topics. To achieve this, KeyATM introduces a new parameter $\pi_{k}$ which represents the probability of sampling from the set of keywords. To train the model, we used all the default settings that as suggested by KeyATM's reference paper. 
 
\noindent\textbf{KeyETM settings:} Like ETM algorithm, we employ variational inference and employ stochastic variational inference
(SVI) to speed up the learning. Both hidden sizes are set to 800, learning rate to 0.005, and the batch size to 40. For 
20Newsgroups dataset based on our discussion in the section 4.1, the $\lambda_{1}=15$ and the $\lambda_{2}=10$, and for the AYLIEN the $\lambda_{1}=25$ and the $\lambda_{2}=10$. 

\subsection{Qualitative Measures}
As the AYLIEN dataset is more complicated, to analyze results qualitatively we used this dataset and compared the learned topics by all models. Table \ref{tab:topwords} displays the top 5 words of 5 topics we defined in Figure \ref{fig:3}, respectively. Topic sequence is matched across topic models and semantically irrelevant words are marked with (X). 

\noindent As shown here, GuidedLDA has the worst performance among others, reporting at least two errors per topic. AVIAD, has the better results compare to GuidedLDA, but it still suffers from irrelevant or overly general words in topics 3, 4 and 5 (e.g., \textit{europ} or \textit{msnbc}). Also compare to KeyATM or KeyETM, AVIAD puts more trivial words in higher ranks (e.g., \textit{suprem} in topic 1 or \textit{nurs} in topic 4). KeyATM, shows impressive performance in topic 1 and topic 4, but it is still unsuccessful in the less dominant topic (BLM). KeyETM shows the best performance among others specifically in the less representative topics. 

In addition to this analysis, we showed in the bottom table in Table \ref{tab:topwords}, the output of the ETM model, as an unguided version of KeyETM model. As you see in this table, two of the topics are mostly related to the election, two of them are about the Covid, and the middle one talks about financial issues. Consequently, the model fails to represent the BLM and social issues topic.
 
\subsection{Measures of Topic Quality}
A good topic model in general should provide both coherent patterns of language and diversity. In other words, the inferred topics not only should be interpretable, but also include a range of aspects. Therefore, we used the following quantitative and behavioral measures to assess topic quality:

\noindent\textbf{Coherence:} Coherence is a widely-used measure of topic interpretability based on the point-wise mutual information between words in each document \cite{lau-etal-2014-machine}. For any two words drawn randomly from the same document, the normalized point-wise mutual information is defined as:
 \begin{equation*}
   f(w_{i},w_{j}) = \frac{log\frac{P(w_{i},w_{j})}{P(w_{i})P(w_{j})}}{-log P(w_{i},w_{j})}
  \end{equation*}
where $P(w_{i},w_{j})$ is the probability of two words co-occurring in a document and $P(w_{i})$ is the marginal probability of word $w_{i}$. Then the topic coherence (TC) is calculated on the top-m (m=10 in our experiments) words as: 
 \begin{equation*}
 TC = \frac{1}{K.\binom{m}{2}}\sum_{k=1}^{K}\sum_{i=1}^{m}\sum_{j=i+1}^{m} f(w_{i}^{(k)},w_{j}^{(k)})
  \end{equation*}
The intuition behind the TC is that the top words of a topic should co-occur often, therefore the higher coherency, the more interpretable topics are.  

\noindent\textbf{Diversity:}
Topic diversity is defined as the percentage of unique words in the top 25 words of all topics \cite{dieng2020topic}. The range is changed between 0 and 1 and higher diversity means more unique words in each topic. As the name implies, the measure captures how semantically diverse the resulting topics are. 

\noindent\textbf{Quality:}
As suggested in \cite{dieng2020topic} the overall metric for measuring quality of a model's topics is calculated as the product of topic diversity and topic coherence. 

\noindent\textbf{Intrusion Task:} We employed a topic intrusion task to assess how human raters evaluated the topic models \cite{chang2009}. The task was administered through Amazon's Mechanical Turk, where eight workers responded to each HIT. Each HIT presented a set of six terms: the five highest probability terms for a given topic, and an intruding term, a low probability term for the given topic but high probability term in one of the other topics. Workers were asked to select the term that did not go with the others. The semantic coherence of the topic is measured by the probability the intruders were successfully identified. To ensure reliable measurement, an intruder was selected from the highly probable words in other topics. As eight workers responded to each hit, we collected a total of 160 intrusion measurements ($5$ topics $\times 4$ intruders $ \times 8$ workers) for each model. As the AYLIEN dataset is more complicated, we applied the Intrusion task just for this dataset. 

\noindent The results for both datasets are shown in Table \ref{tab:TC} and for all three measures (coherence, diversity, quality). They suggest the KeyETM model produces the most coherent and interpretable topics. 

\noindent The result of the intrusion task is also showed in Table \ref{tab:TC} for the four guided topic models. The probability that an intruder was correctly selected (averaged across participants and topics) captures how semantically interpretable the model was to human raters, with a value of $1$ indicating that all subjects correctly identified all the intruders for each topic and a value of $0$ indicating no intruders were correctly identified. We refer to this probability value as the \textit{intrusion score} of the model. For AYLIEN dataset, KeyETM received the highest intrusion score, with a value over $6\%$ higher than the second ranking model. These results suggest that KeyETM produced the most semantically interpretable topics as judged by human raters.

%We fit GuidedLDA, AVIAD, KeyATM(base) and KeyETM with K=5 topics and the seed words sets showed in Figure \ref{fig2}. For each of the four models, we performed quantitative and behavioral assessments of the model. The results are summarized in Table.\ref{eq.4} and KeyETM gives the best performance in terms of all  measures. We further reported an unguided ETM (Labeled ETM) to serve as a reference condition to understand how providing topic models with a small number of keywords substantially improves their performance and better serves the measurement purpose. 

%The results of the intrusion task can be found in Table~\ref{tab:freq}. 
\begin{table}
  \centering
  
  \caption{Measures of Topic Quality.}
  \label{tab:TC}
  \begin{adjustbox}{width=0.5\textwidth}
  \begin{tabular}{l|lcccc}
    \toprule
    &Model&Coherence&Diversity&Quality&Intrusion\\
    \midrule
    \multirow{4}{*}{\textbf{AYLIEN}} &\textbf{GuidedLDA} & 0.03& 0.68 & 0.02 & 0.21\\
    &\textbf{AVIAD} & 0.11 & 0.80 & 0.08 & 0.60\\
    &\textbf{KeyATM(Base)} & 0.14 & 0.75 & 0.11 & 0.69 \\
    &\textbf{KeyETM} & \textbf{0.17} & \textbf{0.86} & \textbf{0.14} &\textbf{0.75}\\
    \midrule
     \multirow{4}{*}{\textbf{20newsg}} &\textbf{GuidedLDA} & 0.06 & 0.76 & 0.04 & -\\
    &\textbf{AVIAD} & 0.17 & 0.81 & 0.13 & -\\
    &\textbf{KeyATM(Base)} & 0.23 & 0.85 & 0.19 & -  \\
    &\textbf{KeyETM} & \textbf{0.25} & \textbf{0.86} & \textbf{0.21} &-\\
    
  \bottomrule
\end{tabular}
\end{adjustbox}
\end{table}
\subsection{Experimental Results for Classification Performance}
Although generative topic models such as KeyETM are proposed as an unsupervised models, but we can also use them to analyze the document representation $\theta_{d}$ and consider its maximum probability as a label. Therefore, the classification performance of all models are evaluated in Table \ref{tab:classify} via precision, recall
and F1 metrics. 

For the AYLIEN dataset, as KeyETM has a better performance to capture the less representative topics (e.g., BLM), it has a better performance to represent documents eventually. As a result, it achieves the highest numbers for all metrics. 
However, for 20Newsgroups dataset, although KeyATM shows slightly better results compared to our proposed model, but as KeyETM uses VAE instead of sampling, the document representations that are generated by the KeyETM are more robust.  
\begin{table}
  \centering
  
  \caption{Measures of Classification.}
  \label{tab:classify}
  \begin{adjustbox}{width=0.5\textwidth}
  \begin{tabular}{l|lcccc}
    \toprule
    &Model&Precision&Recall&F1(Marco)&F1(Micro)\\
    \midrule
    \multirow{4}{*}{\textbf{AYLIEN}} &\textbf{GuidedLDA} & 0.05& 0.06 & 0.05 & 0.05\\
    &\textbf{AVIAD} & 0.19 & 0.19 & 0.17 & 0.18\\
    &\textbf{KeyATM(Base)} & 0.57 & 0.50 & 0.51 & 0.58 \\
    &\textbf{KeyETM} &  \textbf{0.61} & \textbf{0.58} & \textbf{0.57} & \textbf{0.63}\\
    \midrule
     \multirow{4}{*}{\textbf{20newsg}} &\textbf{GuidedLDA} & 0.16& 0.17 & 0.16 & 0.17\\
    &\textbf{AVIAD} & 0.48 & 0.46 & 0.46 & 0.47\\
    &\textbf{KeyATM(Base)} & \textbf{0.87} & \textbf{0.85} & \textbf{0.84} & \textbf{0.85} \\
    &\textbf{KeyETM} & 0.84 & 0.83 & 0.83 &0.83\\
    
  \bottomrule
\end{tabular}
\end{adjustbox}
\end{table}
\section{Discussion and Conclusion}
%\textbf{Labeled ETM} & 0.19 & 0.84 & 0.16 & -\\
 
\begin{comment}
Strikingly, KeyETM had a topic quality score (= coherence $\times$ diversity) over twice as large as the second ranking guided model (see Table~\ref{tab:freq}). 
\end{comment}
Automated topic models tend to retrieve the most general and prominent topics in the corpus as they are not exposed to information regarding the topics of interest. One way to tackle the issue is to incorporate user knowledge in deriving topics by leveraging informative topic-level priors over the vocabulary. Building off the success of ETM which blends the advantages of neural topic models and word embeddings, in this work we propose Keyword Assisted Embedded Topic Model (KeyETM), an embedded topic model guided by a user's knowledge of the corpus.

Through quantitative and qualitative results, we demonstrate that KeyETM retrieves reasonably good topics, relevant to category names we selected earlier. Specifically, it has the best performance in revealing less dominant themes. When compared to baselines which utilize sampling methods, the model has acceptable performance in document representation and classification. This is because it benefits from neural topics modeling and word embeddings. % to derive topics, its results are more robust compared to baselines which utilize sampling methods. 
KeyETM can arm researchers with a tool for extracting desired semantic themes from noisy corpora.

One avenue of future work is to extend KeyETM to enable it to generate robust guided and unguided topics. One approach may be to set the number of topics (K) to a larger value than the number of seed word lists. However, further analysis is needed to validate this approach. %In the current version of KeyETM we can set the number of Topics (K) to a bigger value than the number of seed word sets and the model works properly, but we have not analysis the outcomes to see how the new loss function is worked for this case.  
%%
%% The next two lines define the bibliography style to be used, and
%% the bibliography file.
% \clearpage

\section*{Acknowledgments}
This material is based upon work supported, in part, by the Defense Advanced Research Projects Agency (DARPA) and Army Research Office (ARO) under Contract No. W911NF-21-C-0002.

%\bibliographystyle{ACM-Reference-Format}
%\bibliography{sample-sigconf}
%\input{sample-sigconf.bbl}
%%% -*-BibTeX-*-
%%% Do NOT edit. File created by BibTeX with style
%%% ACM-Reference-Format-Journals [18-Jan-2012].

\end{document}